\documentclass{aa}
\usepackage{graphicx}
\usepackage{natbib}
\bibpunct{(}{)}{;}{a}{}{,}
\newcommand{\be}{\begin{equation}}
\newcommand{\ee}{\end{equation}}
\newcommand{\mytheta}{{\mbox{\boldmath$\vartheta$}}}
\newcommand{\myarcsec}{\hbox{$.\!\!^{\prime\prime}$}}
\newcommand{\myarcmin}{\hbox{$.\!\!^{\prime}$}}

\begin{document}
   \title{A weak lensing analysis of a STIS dark-lens candidate\thanks{based 
          on observations with FORS2@VLT operated by ESO}}

   \author{T. Erben\inst{1}
          \and
          J.M. Miralles\inst{1}
	  \and
	  D. Clowe\inst{1}
	  \and 
	  M. Schirmer\inst{1}
	  \and 
	  P. Schneider\inst{1}
	  \and
	  W. Freudling\inst{2}
	  \and
	  N. Pirzkal\inst{2}
	  \and
	  R. A. E. Fosbury\inst{2}
	  \and
	  B. Jain\inst{3}
          }

   \offprints{T. Erben: terben@astro.uni-bonn.de}

   \institute{$^{1}$Institut f\"ur Astrophysik und Extraterrestrische Forschung (IAEF), 
              Universit\"at Bonn,
              Auf dem H\"ugel 71, D-53121 Bonn, Germany\\
              $^{2}$ST-ECF, European Southern Observatory, Karl-Schwarzschild Str. 2, 
              D-85741 Garching, Germany\\
	      $^{3}$Department of Physics and Astronomy, University of Pennsylvania,
              Philadelphia, PA 19104, USA}

   \date{Received September 15, 1996; accepted March 16, 1997}

   \abstract{We perform a weak lensing analysis on a previously
             reported dark-lens candidate on STIS Parallel data
             \citep{meh02}. New VLT-data indicate that the reported
             signal originates from a small number of galaxies
             tangentially aligned towards the center of the STIS field
             but no signature for an extended mass distribution is
             found.  We argue that we should be able to detect a
             massive cluster ($M\geq 3.2\times 10^{14}M_{\odot}$) 
             through its lensing signal up to a
             redshift of $z\approx 0.6$ with our data.  Also the
             double image hypothesis of two galaxies with very similar
             morphologies in the STIS data is ruled out with colour
             information.   }

   \maketitle
%________________________________________________________________

\section{Introduction}
In the last 5 years, the weak lensing technique has provided us with
important insights into the dark matter distribution of known low- and
high-redshift galaxy clusters [see e.g. \citet{clk00,hfk98,hfk00}].
One of the advantages of this technique over others is that it does
not rely on the relation between the light emission of an object and
its mass, but as \citet{kas93} showed, the surface mass density of
massive objects can be reconstructed from the distortion induced by its
tidal gravitational field on background galaxies.

We can turn around the argument and use this technique to blindly
search for new, hitherto unknown mass concentrations. So far, 9
candidates for such mass concentrations detected by weak lensing
techniques have been reported in the literature
\citep{ewm00,umf00,mwm00,wtm01,meh02,wmt02,dpl02}. Five of them
\citep[][ and two of the three candidates in
\citealt{dpl02}]{wtm01,wmt02,mwm00} have been confirmed as
galaxy clusters.  The other four candidates lack optical counterparts
and are thus potentially dark clusters with an unusually high $M/L$
ratio.  As the firm confirmation of only one such dark cluster would
already have severe consequences for current cluster formation
scenarios and for the nature of dark matter \citep{vpj02, wek02},
careful and independent analyses with follow-up observations have to
be done on such candidates to either reject or strengthen the cluster
hypothesis and to reveal possible, hitherto unknown systematics
in weak lensing analyses.

In this paper we present a weak lensing analysis of ground-based
images of the lens candidate detected in an HST/STIS parallel
field \citep[][ MEH henceforth]{meh02}. The high-resolution STIS image
yielded the strong visual impression of a gravitational lens system.  It
showed several extended objects aligned tangentially to the field centre,
as well as a pair of galaxies with similar morphologies and
surface brightness which seemed to be a potential strong lensing,
double-image candidate, thus indicating lensing by a massive structure.
We show in this work that, despite the first indications, the observed
object configurations in this field most probably do not originate from
gravitational lensing. The paper is organised as follows: First
we present the new
VLT-observations. Thereafter, we make theoretical predictions about
the detectability of a massive galaxy cluster with our data. We
continue by presenting our weak lensing analysis and finish with our
conclusions.
\section{The data}
\label{sec:data}
The VLT-data for the current work were obtained in an ESO Director's
Discretionary Time Proposal (269.A-5064).  The set was observed in the
nights of 05-06/10/2002 with the newly installed FORS2@UT4 camera
having a field-of-view of approximately $7\arcmin\times 7\arcmin$. The
camera consists of two $2\mbox{K}\times 4\mbox{K}$ CCDs with a gap of
about $5\arcsec$ oriented in the West-East direction.  We obtained
3120s $I$-band observations in unbinned mode (pixel scale
$0\myarcsec 126$) as our primary weak lensing science
band. Furthermore, we obtained exposures of 1560s in $V$ and 6380s in
$B$ in the standard binned mode (pixel scale $0\myarcsec 252$) for
detecting a possible red-cluster sequence and to study the
strong-lensing hypothesis of a double-image candidate.  The data in
each band were centered on the position of the STIS field and we
applied a dithering pattern between each individual exposure in order
to minimise the effects of the gap and other defects (hot pixels, bad
columns) of the detectors in the final coadded images.

The data processing was carried out with a pipeline developed specifically
for the reduction of multi-chip cameras that is described in
\citet{ses03}. In the following we only describe the astrometric calibration,
whose accuracy is essential for weak lensing studies, in some more detail.
%For the pre-processing we
%perform standard techniques using the FLIPS software\footnote{see:
%http://www.cfht.hawaii.edu/$\sim$jcc/Flips/flips.html}.  The two chips
%were independently overscan corrected, bias subtracted and flat fielded
%with sky flats where we equalized the gain in the two chips.  From these
%flat fielded science frames, we extracted an illumination correction
%and, for the $I$-band observations, a fringe pattern [fringes are hardly
%visible in the original science frames though, and their amplitudes are
%expected to be below the $1\%$ level (FORS1+2 User Manual)]. After the
%application of the illumination correction and the fringe pattern, sky
%variations in the science frames were smaller than $1\%$ over the
%mosaic.

\begin{figure}
\centering
\resizebox{\hsize}{!}{\includegraphics[width=\textwidth]{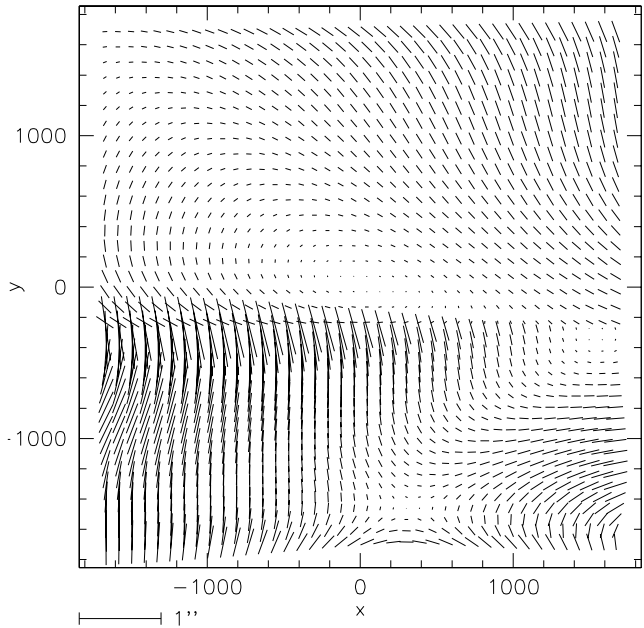}}
   \caption{We illustrate the distortion map from the FORS2 instrument.
            The sticks represent the positional difference of a simple
            shift to match sources in the field with the USNO-A2 catalogue
            compared with a full third order astrometric solution. The
            maximum difference is about 6.19 pixel $(0\myarcsec 78)$.
	    Over a maximum distance of about 2500 pixels from the optical
            axis, the distortion is about $0.247\%$ which is better
            than the specification ($0.3\%$; see also FORS1+2 User Manual).
            The origin of the plot is at the nominal optical axis. We note that
            distortions in the lower chip are significantly higher than
            in the upper one. Correcting for the distortion gives
            us between $1/20-1/10$ pixel internal astrometric 
            accuarcy for overlap objects.}
   \label{fig:distortion}
\end{figure}

After preprocessing, the individual images were astrometrically 
calibrated by comparing object
positions with those from the USNO-A2 astrometric catalogue
\citep{mbc96} having 80 sources in our field.  We used Mario
Radovich's ASTROMETRIX\footnote{see
http://www.na.astro.it/$\sim$radovich/wifix.htm} to fit image
distortions by a third-order, two-dimensional polynomial for every
individual chip. Hereby, distances from objects to the standard star
catalogue and to overlap sources in other chips are minimised in the
$\chi^2$ sense. This procedure gave us positional rms residuals of
$0\myarcsec 15-0\myarcsec 28$ (this is approximately the positional
accuracy of the USNO-A2 catalogue) for the standard stars and
$0\myarcsec 005-0\myarcsec 01$ for the overlap sources in the $I$ band.
Hence, we are able to align the images with an internal accuracy of
1/20-1/10 of a pixel; see also Fig.\ref{fig:distortion}. The final
coaddition was performed with the drizzle package \citep{frh02}. All
the data have been obtained in photometric conditions and photometric
zeropoints were provided by ESO. The measured seeing in the final
coadded $I$ image is about $0\myarcsec 7$.
A first object catalogue was created with SExtractor \citep{bea96},
where we considered all detections with 5 contiguous pixels with
2.5$\sigma$ over the sky background noise. With these parameters we
reached a completeness level of $I\approx 24$ in our primary science 
frame. All magnitudes quoted in this paper are in the Vega system.
\begin{figure}
\centering
\resizebox{\hsize}{!}{\includegraphics[width=\textwidth]{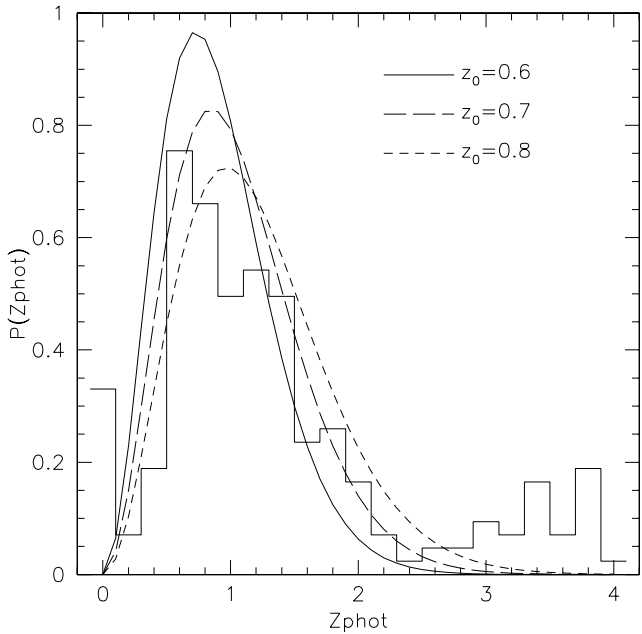}}
   \caption{Shown is the normalised photometric redshift
            distribution $P$(Zphot) around the
            galaxy cluster RXJ1347$-$1145 ($z=0.451$) in $UBVRI$
            photometry. It was estimated with \textit{hyperz}
            \citep[see][ and http://webast.ast.obs-mip.fr/hyperz]{bmp00}. 
            We show only objects with a good redshift
            model fit $P(\chi^2)>90.0\%$ and cut away all galaxies
            having a probability of higher than $60\%$ to be at the
            cluster redshift of $z=0.451$. Although we discovered
            several high redshift candidates in the field, the large
            number of objects having a formal photometric redshift of
            larger than 3 is probably an artefact due to missing
            infrared information.  The three curves are model
            predictions according to eq. \ref{eq:redshiftdistrib} with
            $z_0=0.6$ (solid curve); $z_0=0.7$ (long-dashed curve) and
            $z_0=0.8$ (short-dashed curve).}
   \label{fig:redshift}
\end{figure}
\section{Detectability of massive structures in our data}
\label{sec:detect}
In the following we use standard lensing notation. For
a broader introduction to the topic, see for instance \citet{bas01}.
To estimate the detectability of a galaxy cluster in our current data, we
use the $M_{\rm ap}$ statistic introduced by \citet{sch96_2}. It is
defined as
\be M_{\rm ap}=\int_0^{\theta} {\rm d}^2\mytheta\, \kappa(\mytheta,
   z_{\rm d}, z_{\rm s})\, U(\vartheta), 
\ee 
where
$\int_0^{\theta}d\vartheta\, U(\vartheta)=0$, i.e. $U(\vartheta)$ is a
compensated filter and $z_{\rm d}$, $z_{\rm s}$ are the redshifts
of lens and source, respectively.  Throughout this work we use
$U(\vartheta)=u(\vartheta/\theta)/\theta^2; 
u(\rho)=(9/\pi) (1-\rho^2)(1/3-\rho^2)$.
$M_{\rm ap}$ represents a filtered integral over the surface mass
distribution $\kappa$ and can be related to the tangential shear
$\gamma_{\rm t}$ via
\be
   M_{\rm ap}=\int_0^{\theta} {\rm d}^2\mytheta\, \gamma_{\rm t}(\mytheta)\,
   Q(\vartheta)
\ee
with
\be 
   Q(\vartheta)=\frac 2{\vartheta^2}\int_0^{\vartheta}
   {\rm d}\vartheta' \vartheta' U(\vartheta')-U(\vartheta).  
\ee
For the application to data, it is straightforward to construct an
unbiased estimate $M'_{\rm ap}$ for the integral by a discrete sum
over observed galaxy ellipticities $\epsilon_{\rm t}$ and considering
the coordinate origin being at the center of the aperture:
\be
   M'_{\rm ap}=\frac{\pi\theta^2}N \sum_i\epsilon_{\rm t}(\mytheta_i)Q(\vartheta_i).
\ee
As $M_{\rm ap}$ is a scalar quantity, expectation values for the noise
$\sigma_{\rm Map}$ of a measurement are easily evaluated:
\be
   \label{eq:sigmaMAP}
   \sigma^2_{\rm Map}=\frac{\pi\sigma_{\epsilon}^2}n \int_0^{\theta}{\rm d}\vartheta\,\vartheta Q^2(\vartheta)
   \rightarrow \frac{\sigma_{\epsilon}^2}{2n^2}\sum_i Q^2(\vartheta_i),
\ee
where $\sigma_{\epsilon}$ is the ellipticity dispersion and $n$ 
the number density of galaxies.

For obtaining predictions for expected signal-to-noise $(S/N)$ ratios
for halos with a mass $M$ at a redshift $z_{\rm d}$, we have to
specify a mass model $\kappa(\mytheta, z_{\rm d}, z_{\rm s})$ and a
distribution for the source redshifts $p(z_{\rm s})$. The expected
signal $S$ is then given by
\be
   S=\int_0^{\theta}{\rm d}^2\mytheta\int_{z_{\rm d}}^{\infty} {\rm d}z_{\rm s}\,
   \kappa(\mytheta, z_{\rm d}, z_{\rm s})U(\vartheta)p(z_{\rm s}).
\ee
The noise is given by $\sigma_{\rm Map}$ in (\ref{eq:sigmaMAP}). For
the mass profile we consider the universal density profile proposed by
\citet{nfw96}. The details, how the surface mass density
$\kappa(\mytheta, z_{\rm d}, z_{\rm s})$ is obtained for this profile
when fixing the halo mass $M$, are given in \citep{bar96,krs99} and will not
be repeated here.  For the source redshifts we consider the normalised
distribution:
\be
   \label{eq:redshiftdistrib}
   p(z_{\rm s})=\frac 3{2z_0}\left(\frac {z_{\rm s}}{z_0}\right)^2
   \exp\left[ -\left(\frac {z_{\rm s}}{z_0}\right)^{1.5}\right]
\ee
which was proposed in \citet{bbs96}. To fix $z_0$, we considered the
redshift distribution from $UBVRI$ photometry around the galaxy
cluster RXJ1347$-$1145. The data for this cluster will be described
elsewhere (Erben et al., in preparation). They have been obtained with
FORS1 around the cluster center (field-of-view $3\arcmin\times
3\arcmin$). As for the current data set, the $I$ band of these observations
is used as the primary science band for a weak lensing study of the cluster.
As the two observations reach about the same depth ($I\approx 24$), we expect
that the source galaxies trace comparable redshift distributions.
The redshift distribution for the cluster field is shown in
Fig. \ref{fig:redshift}. We estimate from this figure that $z_0=0.7$
(and thus $\langle z_s\rangle=1.05$)
provides a fair redshift description for the source distribution 
in our current data. As the lens magnification effect of the 
massive cluster might significantly push the redshift distribution
to higher values we also investigate the case of $z_0=0.6$;
$\langle z_s\rangle=0.9$.

\begin{figure}
\centering
\resizebox{\hsize}{!}{\includegraphics[width=\textwidth]{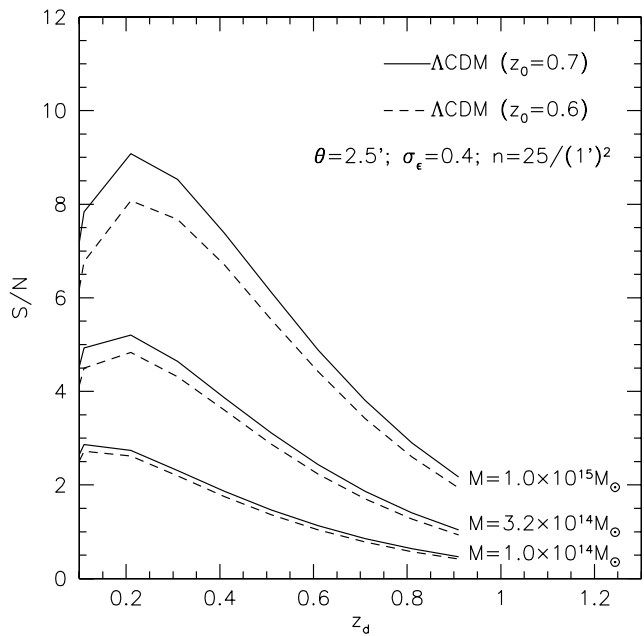}}
   \caption{Shown are $S/N$ predictions for the detectability of
            massive structures in our VLT data. We plot predictions
            for a $\Lambda$CDM cosmology ($\Omega_0=0.3; \Omega_{\Lambda}=0.7;
            \sigma_8=0.9; \Gamma=0.25$) and two different
            redshift distributions (see eq. \ref{eq:redshiftdistrib}).
            We assumed a smoothing
            scale of $\theta=2\myarcmin 5$ [\citet{krs99} showed that
            for a smoothing scale of about $2\arcmin$, $M_{\rm ap}$
            gives the maximum significance for halos following a NFW
            profile], a number density of 25 galaxies per sq.
            arcminute and $\sigma_{\epsilon}=0.4$ for the width of the
            ellipticity distribution (measured to be
            $\sigma_{\epsilon}=0.42$ in our data). The results do not change
            significantly if we consider an EdS universe 
            ($\Omega_0=1.0; \Omega_{\Lambda}=0.0;\sigma_8=0.6; \Gamma=0.25$) 
            instead.}
   \label{fig:clumpSN}
\end{figure}

Fig. \ref{fig:clumpSN} shows our predictions for $S/N$ ratios from the
described model.  We conclude that we are insensitive to structures
below $10^{14}M_{\odot}$ and can marginally detect, at the $3\sigma$
level, clusters of about $3\times 10^{14}M_{\odot}$ up to a redshift of
$z\approx 0.6$.
\section{Weak lensing analysis of the VLT-data} 
\label{sec:wl}
To cross-check our weak lensing results, we performed two independent
analyses of the data.\\ \\ 
{\bf Analysis 1:}\\ 
Starting from the
initial SExtractor catalogue we determined, for all the objects, the
quantities necessary to obtain shear estimates according to the KSB
\citep{ksb95} algorithm. To correct all galaxies for PSF anisotropy
and PSF smearing effects we closely follow the procedures described in
\citet{ewb01}.  There we introduced a weighting scheme for individual
galaxies based on the ellipticity distribution of the corrected
ellipticities. Since in the current paper we were looking for a 
potential strong lensing
cluster, high ellipticities could be caused by lensing and do not
necessarily reflect the true noise properties. Hence, we did not use
this weighting scheme for the current work but conservatively rejected
all galaxies having a corrected ellipticity larger than 0.8. The final
catalogue for the lensing analysis contains 1200 objects, i.e. around
25 per square arcmin.

\begin{figure}
\centering
\resizebox{\hsize}{!}{\includegraphics[width=\textwidth,angle=-90]{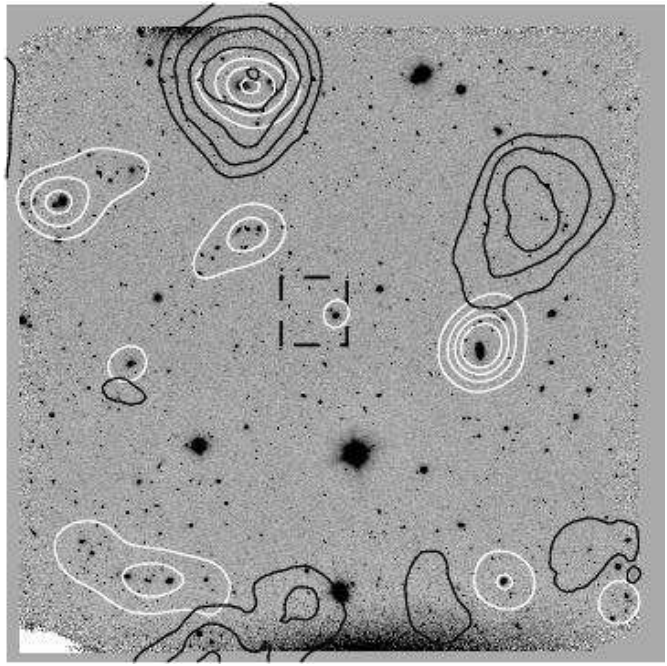}}
   \caption{Shown is the light distribution (white contours) and
   $M_{\rm ap}$ significance contours (black contours) in our VLT
   field. The light distribution contains all galaxies also used
   for the weak lensing analysis. It is estimated on a regular grid,
   where each gridpoint contains the total flux within $1\arcmin$
   weighted by a Gaussian with a width of $22\myarcsec 5$.
   The light contours enclose regions where the light is 3, 6,
   9 and $12\sigma$ above the mean light. The $M_{\rm ap}$ contours
   display mass significance of 1.0, 1.5, 2.0, 2.5 and 3.0$\sigma$
   with a smoothing scale of $2\myarcmin 5$.  The dashed rectangle marks
   the area of the original HST/STIS observation. We see that no
   significant detection is found at the position of our dark lens
   candidate.}
   \label{fig:map_1200}
\end{figure}

\begin{figure}
\centering
\resizebox{\hsize}{!}{\includegraphics[width=\textwidth,angle=-90]{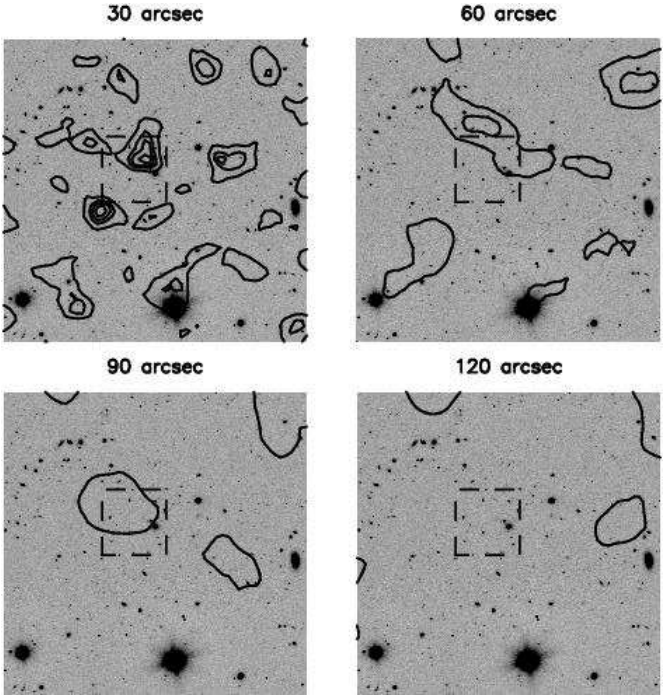}}
   \caption{Shown are $M_{\rm ap}$ significance contours for different filter
   scales indicated by the labels. Contours show the levels of $1.0, 2.0, 2.5$ and
   $3.0\sigma$. Dashed rectangles mark the original STIS observation.
   We note that the $3\sigma$ peak at the position of
   our lens candidate at the smallest filter scale vanishes
   when the filter size is increased.}
   \label{fig:sn_smallfield}
\end{figure}

Figs. \ref{fig:map_1200} and \ref{fig:sn_smallfield} show the $M_{\rm
ap}$ statistics result for our data. As can be seen we do not recover
any significant mass concentration at the position of our original
lens candidate when the filter scale becomes larger than the size of
the original STIS image. We can recover a $3\sigma$ peak with a filter
scale of $30\arcsec$ which is consistent with the $2.5\sigma$ detection
reported on the original STIS data. The light distribution also shows no
overdensity in the region under consideration. 
\newline
\newline
{\bf Analysis 2:}\\ 
For the second analysis, we used SExtractor
to create a catalogue of all objects in the $I$-band image which had at
least three pixels with flux greater than the $1\sigma$ sky level
noise.  This catalogue, which contained mostly noise peaks, was then
analysed by convolving each object in the image with progressively
larger Mexican-hat filters until a maximum in signal-to-noise $\nu$,
for the filtered object was found.  The background galaxy catalogue was
then selected from objects with SExtractor isophotal magnitude $I >
22$, $\nu > 10$, having a Mexican-hat filter radius at maximum $\nu$
which was larger than that measured for stars, and not having a
neighboring object within $2\farcs 5$.  The resulting catalogue
contained 1838 objects, with 954 in common with the catalogue from the
first analysis.  Objects were detected as faint as $I=27$, although
the completeness limit, as judged by where the number counts depart
from a powerlaw, is at $I\sim 24.8$.

The background galaxy ellipticities were also corrected for PSF smearing to
obtain shear estimates using the KSB formalism, but the methods
used to reduce the noise in the shear and smear polarizabilities were
different from those employed in the first analysis.  The method was
that given in \citet{cls01}, using fifth-order
two-dimensional polynomials to fit the stellar ellipticity and shear
and smear polarizabilites as a function of position in the image.  A
direct comparison of the catalogues for the two analyses results in the
same mean shear across the objects and an rms shear difference of
$0.16$ for the objects in common, while the rms shear of the objects
is about $0.4$.  

\begin{figure}
\centering
\resizebox{\hsize}{!}{\includegraphics[width=\textwidth,angle=-90]{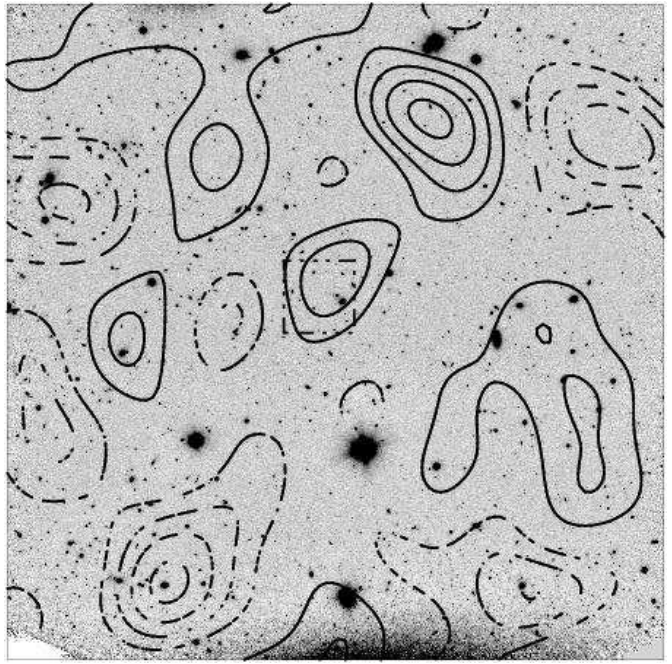}}
   \caption{The figure shows the weak lensing mass reconstruction
   \citep{kas93} from the second analysis. The smoothing of the Map is $28\arcsec$.
   The highest peak (roughly on the $2.5\sigma$ level) is close to the
   second peak in the North-West of the $M_{\rm ap}$ map from TE
   (Fig. \ref{fig:map_1200}).  No other features above the noise level
   are seen in this reconstruction. Dashed contours show regions of
   negative $\kappa$. The contours mark 23\%, 46\%, 69\% and 92\% of the
   maximum and minimum peak intensity in the image respectively.}
   \label{fig:doug_map}
\end{figure}

A mass reconstruction from direct Fourier transform of the shear field
\citep{kas93} is shown in Fig. \ref{fig:doug_map}.  The large peak in
the NNW portion of the image is signficant at $2.5\sigma$, while all
of the other peaks are less than $2\sigma$.  There is a detected
overdensity in the vicinity of the dark lens candidate, but at only a
$1\sigma$ significance. In the next section we analyse the implications 
of the current null result by taking a closer look at the galaxy 
populations used in this and the original analysis.
\section{Interpretation}
The original STIS data gave a striking visual impression of a strong
gravitational lens manifesting itself in several highly elongated
arclet candidates showing a coherent alignment towards the brightest
galaxy in the field (see Fig. 2 of MEH and the publicly available data
under http://www.stecf.org/projects/shear/slens/slens.fits).  If the
features present in this field indeed originate from gravitational
lensing and hence from a mass concentration of the order of a massive
galaxy cluster, we should expect to detect a significant $M_{\rm ap}$
signal with a filter scale of about $2\arcmin$ (see
Sect. \ref{sec:detect}) or a significant peak in a weak lensing mass
reconstruction. Also, a larger scale overdensity in the light
distribution should be detected for a massive galaxy cluster with a
typical mass-to-light ratio. With the available VLT data covering the
surroundings of the candidate on a field of about $7\arcmin\times
7\arcmin$, we fail to find both signatures of a possible massive
structure at the STIS image position.

As was shown in Sect. \ref{sec:detect}, we should be able to detect
either a massive galaxy cluster up to a redshift of $z\approx 0.6$
with the data at hand, or a medium-sized cluster up to a redshift of
$z\approx 0.3$. 

Hence, upholding the lensing hypothesis would imply that the
massive structure is at substantially higher redshift that we cannot
probe with our VLT data. Since we reach a significantly deeper source
population with our STIS data (see also Fig. \ref{fig:hst_vlt}), we
have to investigate this possibility further. In the following we try
to test this hypothesis by considering common objects in the STIS and
VLT analysis and their photometric properties.  In the original
analysis we split the detected sources in the STIS image into two
samples. The first sample consists of 52 objects which represents
the catalogue for our weak lensing analysis using the $M_{\rm ap}$
statistics. This analysis yields a peak at the $2.4\sigma$
level. In the following, we regard the position of this peak as the
center of our lens candidate.  A second, independent object sample
with 11 objects consists of all highly elongated and well resolved
objects having a SExtractor axis ratio larger than 2. Most of them were
rejected from the weak lensing catalogue for various reasons (for
instance because of difficulties in determining an exact object center
or having a too large corrected ellipticity) and we regard them as
lensed arclet candidates. Considering the position angle distribution,
we found that the probability for the observed tangential alignment of
these objects with respect to the $M_{\rm ap}$ center was only $0.3\%$
when assuming a random angle distribution. In this way we ended up
with two independent lensing signatures of a massive object at the
2.5$\sigma$ level. Moreover, two of these arclet candidates were
considered as a multiple images candidate because of their
morphological similarity and nearly identical surface brightness in
the STIS Clear filter. On these issues, our VLT data shed new,
important light.
 
\begin{figure}
\centering
\resizebox{\hsize}{!}{\includegraphics[width=\textwidth,angle=-90]{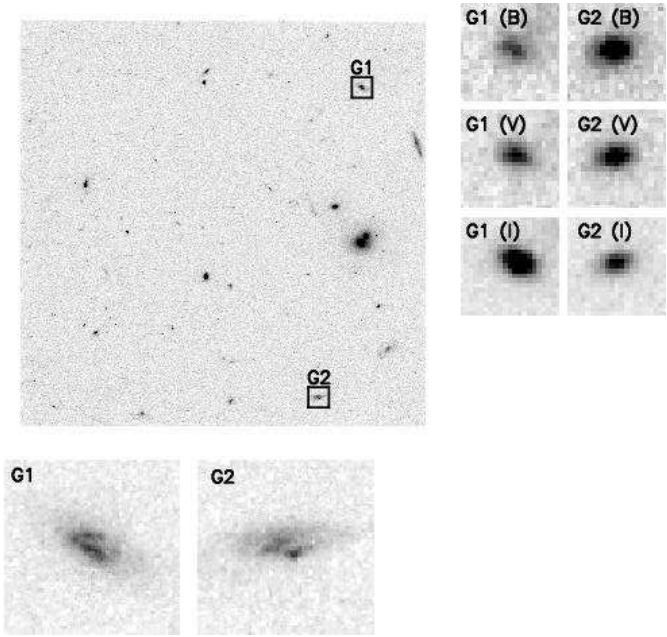}}
   \caption{We show the status of the double image candidate G1
   and G2 presented in MEH. The surface brightness for the two objects
   in the STIS Clear filter is nearly identical (23.29 mag/arcsec$^2$
   for G1 and 23.34 mag/arcsec$^2$ for G2) and also the very similar
   and striking morphology suggested them to be two images of a single
   source. Equal parity of the images weakened the lensing
   hypothesis, though. Regarding the brightness in the three VLT
   colours (panels on the right) immediately reveals very different
   object colours ($B-I$=3.85 for G1 and $B-I$=2.42 for G2), excluding
   the lensing hypothesis.  }
   \label{fig:g1g2}
\end{figure}

\begin{figure*}
\centering
\resizebox{\hsize}{!}{\includegraphics[width=\textwidth,angle=-90]{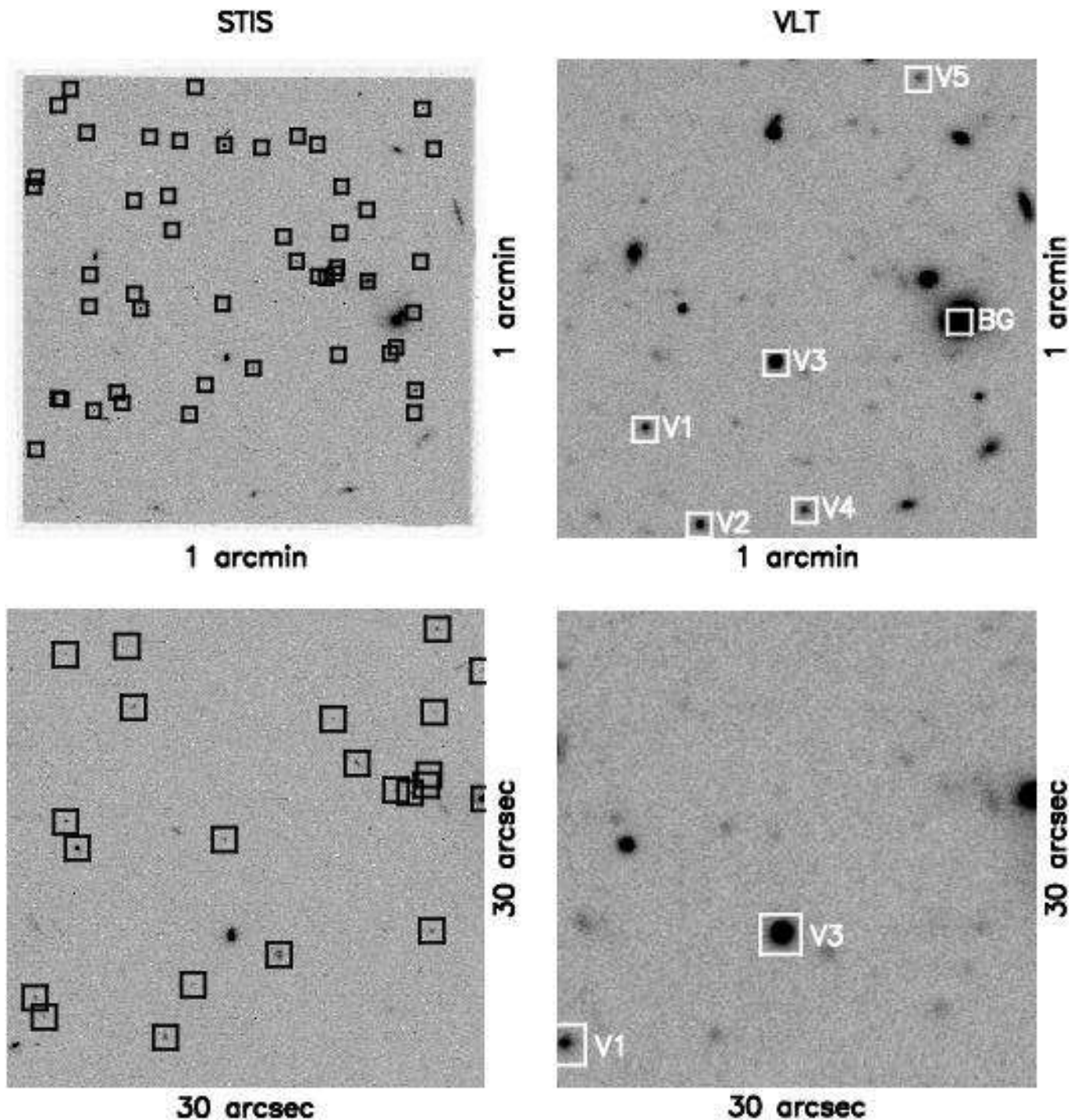}}
   \caption{Shown are common objects in the original STIS and the
   current VLT weak lensing work (white boxes in the upper right panel;
   the plot does not include common objects that were amongst the STIS
   strong lensed arclet candidates), objects that
   were used in the STIS analysis only (black boxes in the upper left
   panel). The lower panels are $30\arcsec$ zooms from the upper
   panels. We see that many galaxies that were used in the HST/STIS
   analysis are still visible but too faint or too small to allow
   a reliable shape measurement in our high-quality VLT observations.}
   \label{fig:hst_vlt}
\end{figure*}

First, they reveal that the two objects of our double image candidate
(G1 and G2 of Fig. 2 in MEH) are different sources and cannot be due
to a single, lensed object because of their colours.  Object G1 has
$B-V=1.53$ and $V-I=2.32$, while G2 has $B-V=1.12$ and
$V-I=1.30$. Fig. \ref{fig:g1g2} summarises the situation for these two
objects.

Next, we investigate in more detail whether the candidate arclets can
still be considered as such. If their ellipticities are indeed caused
by a massive structure at high redshift ($z\ge 0.6$) they must be
located at even substantially higher redshift.  The three bands in
hand do not permit the estimate of accurate photometric
redshifts \citep{bmp00} for these objects, but do permit us to test
whether high redshifts (larger than $z\approx 1$) would be compatible
with their colours. Adding a prior about the luminosity function of
the galaxies helps to constrain the permitted redshift ranges
with only three bands [see e.g. \citet{ben00}]. We choose to
consider only solutions for photometric redshifts which are compatible with an absolute
magnitude range $M_* - 2 < M < M_* + 3$ for the $I$-band. This band is
rather insensitive to evolution effects since it probes the old
stellar population, and we choose to use the local $M_*=-21.26$
\citep{bde00} deduced from the SDSS early data release.  In Table
\ref{table:redshiftranges} we present our deduced redshift constraints
for the original arclet candidates (objects G1, G2 and A1-A8 in Fig. 2
of MEH).
\begin{table*}
\begin{center}
\caption{Given are the photometric redshift intervals (estimated with
{\sl hyperz}) at the 68\% confidence level using the $B,V$ and $I$
images.  Peaks in the probability distribution are also given.  We
determine only regions so that the absolute $I$ magnitudes of objects
lie within $M_* - 2 < M < M_* + 3$ as explained in the text. We
consider the original strong lensed arclet candidates (G1, G2 and
A1-A8), the brightest galaxy in the STIS field (BG) and 5 additional
objects (V1-V5) common to the STIS and VLT weak lensing analysis.  See also
Fig. \ref{fig:hst_vlt}.}
\begin{tabular}{|c|c|c|c|c|}
\hline
object & $I-$mag & permitted redshift range & peaks & comments \\
\hline
{\bf G1} & 21.74 & $0.45 < z < 0.75$ & 0.48; 0.67 & \\
{\bf G2} & 22.54 & $0.3 < z < 0.5$ & 0.34; 0.45 & \\
         & & $1.62< z < 1.66$ & 1.64 & \\
{\bf A1} & 21.44 & $0.49< z < 0.75$ & 0.54; 0.64 & \\
{\bf A2} & 21.55 & $0.34 < z < 0.415$ & 0.35 & two resolved objects\\
         & & $0.48< z <0.52$ & 0.51 & in STIS \\
{\bf A3} & - & - & - & merged with a star in VLT data \\
{\bf A4} & 24.34 & $0.5< z < 1.6$ & 0.59; 0.82; 1.37 & \\
{\bf A5} & 25.10 & $1.49< z <1.79$ & 1.64 & \\
{\bf A6} & 24.37 & $1.49< z <1.75$ & 1.55 & \\
{\bf A7} & - & - & - & not detected in VLT images \\
{\bf A8} & 25.05 & $1.6< z <2.7$ & - & multiple peaks \\
{\bf BG} & 19.09 & $0.29 < z < 0.43$ & 0.35 & \\
\hline
{\bf V1} & 23.20 & $0.45< z <0.58$ & 0.5 & \\
           & & $1.32< z <1.98$ & 1.48 & \\
{\bf V2} & 22.80 & $0.57< z <1.01$ & 0.71; 0.86; 0.98 & \\
           & & $1.28< z <1.45$ & 1.30; 1.41 & \\
{\bf V3} & 21.66 & $0.42< z <0.56$ & 0.44; 0.54 & \\
{\bf V4} & 23.32 & $0.42< z <0.46$ & 0.45 & \\
           & & $1.64< z <1.69$ & 1.67 & \\
{\bf V5} & 23.86 & $1.83< z <1.87$ & 1.85 & marginal solutions \\
           & & & & at $z\sim$0.3 and $z\sim$3 \\
\hline
\end{tabular}
\label{table:redshiftranges}
\end{center}
\end{table*}
We note that the colours of the brightest and most prominent
candidates (G1, G2, A1, A2 and A4) are compatible mostly with redshifts
considerably lower than unity. All the other faint objects
in the sample have indeed a high probability of being placed at $z\geq
1.4$.

However, with the most prominent arclet candidates being probably at low
redshift and without the multiple image candidate, the strong lensing
hypothesis for these fainter sources by a high-redshift cluster looses
much of its strength.

Finally, we take a closer look at the original weak lensing signal of
our STIS field.  The $2.4\sigma$ $M_{\rm ap}$ signal has been very
robust against the analysis of various subsamples of the 52 objects. We
still could recover peaks close to the original signal when removing
randomly either 30\% and 50\% of the sources or repeating the
procedure with the most elliptical sources.  Fig. \ref{fig:hst_vlt}
shows that only 5 of these objects that are also amongst the brightest
in the whole STIS field are also used in the first VLT analysis (not
considering the arclet candidates). Considering the allowed redshift
ranges as we did with the arclet candidates (see Table
\ref{table:redshiftranges}) indicates that these sources probably lie
at low redshift. These sources lie at the faint end of the VLT sources
but only at the bright end of the STIS galaxy population used in the
initial analysis.  Hence, as we are not able to use most of the
original, faint STIS galaxy population in our current weak lensing
analysis, no conclusive answers on the origin of the initial weak
lensing signal can be given at this stage.
\section{Conclusions and Outlook}
\label{sec:conclusions}
In MEH we reported on the conspicious tangential alignment of 63
galaxies in a $52\arcsec\times 52\arcsec$ high resolution STIS
field. The maximum of the weak lensing $M_{\rm ap}$ 
signal of 52 objects
defined a formal mass peak close to the brightest elliptical galaxy in
the field. Eleven other independent and elongated objects show a
coherent tangential alignment towards the center defined by $M_{\rm
ap}$. In addition, two of those objects have very similar morphologies
and surface brightnesses in the original STIS data suggesting a multiple
lensed image. Hence, the
discovery of a new, massive galaxy cluster acting as a strong
gravitational lens had to be taken seriously into account and to be
further investigated with optical follow-up observations covering a larger
field-of-view. With new, high-quality VLT data in three bands covering
$7\arcmin\times 7\arcmin$ around the original cluster candidate, we
now arrive at the following conclusions:
\begin{itemize}
\item The light distribution in the VLT field shows that the
brightest galaxy in the STIS field is an isolated early type
galaxy and not part of an optically rich galaxy cluster. There is no
indication for an overdensity of the light distribution in or very
close to the STIS field.
\item With the data at hand we should detect, with $\geq 3\sigma$
significance, a matter concentration of about $3\times
10^{14}M_{\odot}$ up to a redshift of $z\leq 0.6$ with weak lensing
techniques. We fail to find such a lensing signature in two
independent analyses.
\item We can exclude the initial double image hypothesis for
two of the very elongated objects we regarded as strongly lensed
arclet candidates. Moreover, the colours of the brightest and most 
prominent of these candidates are only compatible with redshifts
up to about 0.7. This makes it impossible that their large elongations
originate from lensing by a high-redshift cluster.
\end{itemize}
Regarding the origin of the original weak lensing signal in the
STIS data, no progress could be made with the new VLT data at
hand. Besides the brighter arclet candidates, only 5 additional
objects are used in common in the original STIS and the new VLT weak
lensing analysis. Those represent the faint end of the VLT but only
the bright end in the STIS galaxy population. The remaining 47 STIS
sources are too small and too faint in the high-quality VLT
images to be usable for a weak lensing study. Hence, we cannot exclude
that those sources represent a high redshift galaxy population and
their very robust alignment is indeed caused by lensing by a mass
concentration at high redshift.

With the negative conclusions regarding the initial strong lensing
arguments, especially the loss of the double image candidate, a
chance alignment of 52 objects on a $52''\times 52''$ field causing a signal
on the $2.5\sigma$ level is the most plausible explanation for the
initial signal at this point.

The past has shown that deep HST exposures are likely to reveal
highly elongated and well resolved galaxies with lens-like morphologies. 
Also, for the Hubble Deep Field North, a possible strong
lensing configuration has been found \citep{hbk96} but the lensing
hypothesis has later been falsified with Keck spectroscopy 
\citep{zmd97}. Most of these strong-lens candidates found in HST data
can be resolved by follow-up observations on ground-based facilities,
either by spectroscopy or multi-colour observations. 

In contrast to clarifying these strong lens candidates, it is more
difficult to confirm or discard potential discoveries of mass
concentrations found with weak lensing techniques on HST images on
small angular scales, as in the current case, or that presented in
\citet{umf00}. The small field-of-view and limited number of galaxies
increase the probability for chance-alignments compared with
wide-field ground-based observations. On the other hand, as was shown
by this and previous studies [see e.g. \citet{hfk00}], current state
of the art ground based facilities are not able to probe the faint
galaxy population of deep HST exposures with a resolution sufficient for
weak lensing studies. Currently, only new space-based observations
with ACS, having a large enough field-of-view to probe a sufficiently
large area around these candidates, could finally close these cases.
\begin{acknowledgements}
We thank the ESO Director General, C. Cesarsky for approval of the
DDT proposal and the service observing team at Paranal for providing
us with this excellent data set. We are grateful to L. King and
to the referee, J. D. Rhodes for a careful reading of the manuscript. 
This work was supported by the German Ministry for Science
and Education (BMBF) through the DLR under the project 50 OR 0106, by
the German Ministry for Science and Education (BMBF) through DESY
under the project 05AE2PDA/8, and by the Deutsche
Forschungsgemeinschaft (DFG) under the project SCHN 342/3--1.
\end{acknowledgements}
\bibliographystyle{aa}
\bibliography{slens}
\end{document}